\documentclass{PoS}

\usepackage{epsfig}

\def\Journal#1#2#3#4{{#1} {\bf #2} (#3) #4}

\def\PLB{{\em Phys. Lett.}  B}

\def\PRD{{\em Phys. Rev.} D}

\def\EPJC{{\em Eur. Phys. J.} C}

\def\PRT{\em Phys. Rept.}

\def\etal{{\it et al.}}

\def\NPPS{\em Nucl.\ Phys.\ Proc.\ Suppl.}

\def\be{\begin{equation}}
\def\ee{\end{equation}}
\def\bea{\begin{eqnarray}}
\def\eea{\end{eqnarray}}

\newcommand{\dd}       {\mathrm{d}}

\newcommand{\bbbar}     {\ensuremath{\mathrm{b\bar{b}}}}

\newcommand{\epem}              {\ensuremath{\mathrm{e^+e^-}}}

\newcommand{\as}                {\ensuremath{\alpha_\mathrm{S}}}

\newcommand{\asmz}              {\ensuremath{\alpha_\mathrm{S}(M_{\mathrm{Z^0}})}}

\newcommand{\momone}[1] {\mbox{\ensuremath{\langle#1\rangle}}}
\newcommand{\momn}[2] {\mbox{\ensuremath{\langle#1^{#2}\rangle}}}
\newcommand{\cp}                {\ensuremath{C}}
\newcommand{\bt}                {\ensuremath{B_\mathrm{T}}}
\newcommand{\bw}                {\ensuremath{B_\mathrm{W}}}
\newcommand{\mh}                {\ensuremath{M_\mathrm{H}}}

\newcommand{\thr}               {\ensuremath{1-T}}
\newcommand{\ytwothree}         {\ensuremath{y_{23}}}

\newcommand{\chisqd}    {\ensuremath{\chi^2/\mathrm{d.o.f.}}}
\newcommand{\xmu}               {\ensuremath{x_{\mu}}}

\newcommand{\ycut}              {\ensuremath{y_{\mathrm{cut}}}}

\newcommand{\stat}              {\ensuremath{\mathrm{(stat.)}}}

\newcommand{\expt}               {\ensuremath{\mathrm{(exp.)}}}
\newcommand{\had}               {\ensuremath{\mathrm{(had.)}}}
\newcommand{\theo}              {\ensuremath{\mathrm{(theo.)}}}
\newcommand{\tot}              {\ensuremath{\mathrm{(tot.)}}}

\newcommand{\rs}                {\ensuremath{\sqrt{s}}}

\newcommand{\invpb}     {\ensuremath{\mathrm{pb}^{-1}}}

\newcommand{\py}                {PYTHIA}
\newcommand{\hw}                {HERWIG}
\newcommand{\ar}                {ARIADNE}

\newcommand{\debr}    {DEBRECEN 2.0}

\newcommand{\resultjr} {
\ensuremath{\asmz=0.1169\pm0.0004\stat\pm0.0012\expt\pm0.0021\had\pm0.0007\theo}}
\newcommand{\resultjrtot} {
\ensuremath{\asmz=0.1169\pm0.0026\tot}}

\newcommand{\resultmom} {
\ensuremath{\asmz=0.1286\pm0.0007\stat\pm0.0011\expt\pm0.0022\had\pm0.0068\theo}}
\newcommand{\resultmomtot} {
\ensuremath{\asmz=0.1286\pm0.0072\tot}}

\title{Measurement of \as\ with JADE using Moments
of event shape Observables and the Four-Jet Rate}

\ShortTitle{\as\ with JADE using event shape moments and the four-jet rate}

\author{\speaker{Jochen Schieck}\\
        Max-Planck-Institut f\"ur Physik\\
	Munich, Germany\\
        E-mail: \email{schieck@mppmu.mpg.de}}
\abstract{
  Data from \epem\ annihilation into hadrons collected by the JADE
  experiment at centre-of-mass energies between 14~GeV and 44~GeV were
  used to study moments of event shape distributions and the four-jet 
  rate as a function of the Durham algorithm's resolution 
  parameter \ycut. The data were compared to a QCD NLO order 
  calculations and including NLLA resummation in the case
  of the four-jet rate. The strong coupling measured 
  from the moments was
\begin{center}
\resultmom, 
\end{center}
and from the four-jet rate was
\begin{center}
\resultjr, 
\end{center}
both in agreement with the world average.}
\FullConference{International Europhysics Conference on High Energy Physics\\
                 July 21st - 27th 2005\\
                 Lisboa, Portugal}
\begin{document}
\section{Introduction}
The annihilation of an electron and a positron into hadrons allows
precise tests of Quantum Chromodynamics (QCD). 
We measured the first five moments of event shape 
observables and the four-jet rate in \epem\ annihilation 
and compared the data to predictions by Monte Carlo Models and by perturbative
QCD. From the comparison of the data with the theory we extracted the
strong coupling \as. 
A more detailed description of this analysis can be found in ~\cite{jadenote}. 
We used data collected by the JADE experiment~\cite{naroska87} in the
years 1979 to 1986 at the PETRA \epem\ collider at DESY at six
center-of-mass energies of $\rs=14.0, 22.0, 34.6, 35.0, 38.3$ and 
$43.8$~GeV. The total integrated luminosity was 195 \invpb\ corresponding
to about 40000 events. Large samples of Monte Carlo simulated events were 
used to correct the data
for experimental acceptance, resolution, backgrounds and fragmentation effects.
The process $\epem\to\mathrm{hadrons}$ was simulated using \py\ and 
corresponding samples using \hw\ and \ar\ were used for 
systematic checks. 
The event selection for this analysis aimed to identify hadronic
event candidates and rejected events with a large amount of energy
emitted by initial state radiation (ISR), leptonic final states and two photon
events.
Events from the process $\epem\rightarrow\bbbar$ were
considered as background, since especially at low center-of-mass energy the large
mass of the b quarks and of the subsequently produced B hadrons will
influence the measurement. Therefore the contribution from
expected \bbbar\ events was subtracted ~\cite{jadenewas}.
The effects of detector acceptance and resolution and
of residual ISR were accounted for by a multiplicative correction
procedure. 
In order to confront the theoretical prediction valid for partons
with the data the theoretical prediction was corrected 
for hadronization with \py.
Several sources of possible systematic uncertainties were studied.
The experimental uncertainties evaluate deficiencies in the 
reconstruction of the data and the modelling of the detector. The
uncertainties associated with the correction for hadronization
effects were assessed by using alternative models (\hw\ and \ar\ instead
of the default \py). The theoretical uncertainty associated with 
missing higher order terms in the theoretical prediction, was assessed 
by setting the renormalization scale \xmu\ to 0.5 and 2 instead of the default value
one.
The results from the fits at the various center-of-mass energies were
combined in order to determine a single value of \asmz.
\section{Event Shape moments}
\begin{figure}
\centerline{
\includegraphics[width=.5\textwidth]{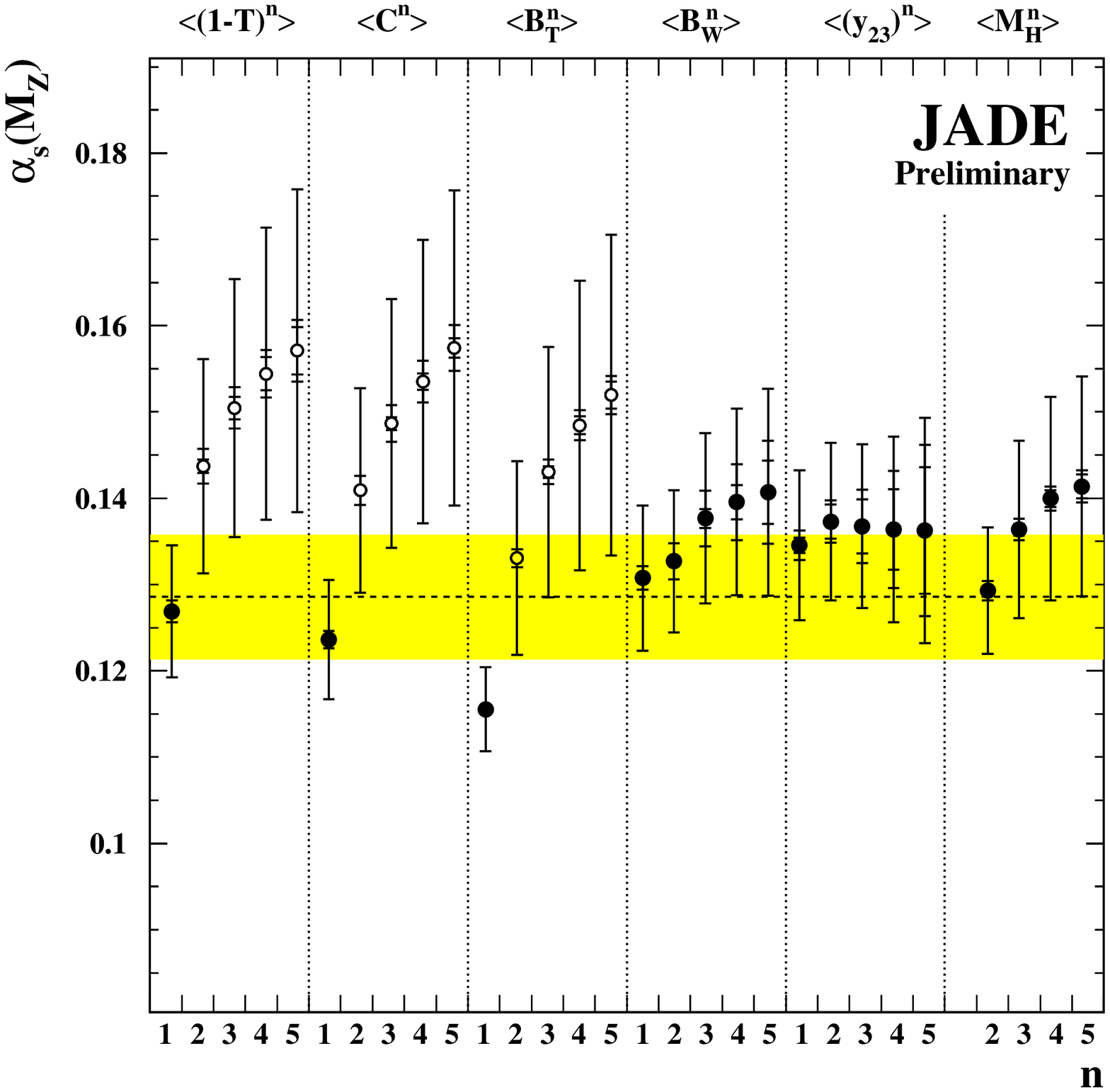}
\includegraphics[width=.5\textwidth]{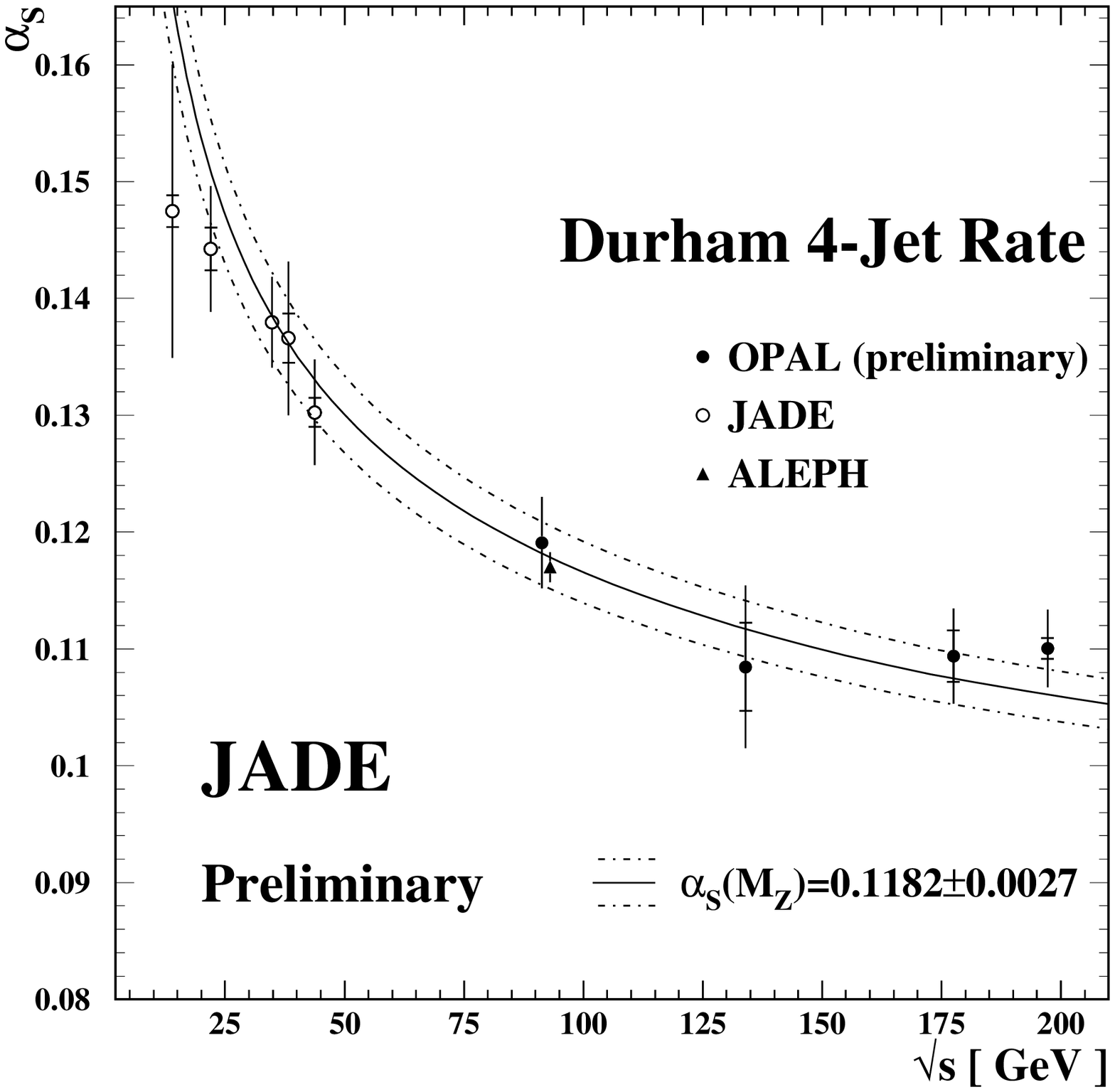}
 }
\caption{
The figure on the left shows the measurements of \asmz\  using fits 
to moments of six event shape observables.  The inner error bars 
represent statistical errors, the middle error bars include experimental 
errors and the outer error bars show the total errors.  The dotted line indicates
  the weighted average described in the text; only the measurements
  indicated by solid symbols were used for this purpose.
The figure on the right shows the values for \as\ using the four-jet rate at 
  the various energy points. The errors
  show the statistical (inner part) and the total error.  The full and
  dash-dotted lines indicate the current world average value of
  \asmz~\cite{bethke04}.  The results at $\rs=34.6$ and 35~GeV have
  been combined for clarity.  The results from ALEPH~\cite{aleph249}
  and OPAL~\cite{OPALPN527} (preliminary) are shown as well.
\label{alphas}
}
\end{figure}

The properties of hadronic events may be characterized by 
event shape observables. The following event shapes 
were considered in this analysis: 
Thrust, C-Parameter, Heavy Jet Mass, Jet Broadening
Observables \bt\ and \bw\ and the transition 
value between 2 and 3 jets \ytwothree\ using the Durham scheme.
The $n$th, $n=1,2,\ldots,5$ moment of the distribution of an event
shape observable $y$ was defined by
$\momn{y}{n}=\int_0^{y_{max}} y^n \frac{1}{\sigma}
\frac{\dd\sigma}{\dd y} \dd y \;\;\;,$
where $y_{max}$ was the
kinematically allowed upper limit of the observable.
The measurement involved a full integration over the available phase space.
Comparisons of QCD predictions were thus complementary
to tests of the theory using the differential distributions. 
In the case of the moments of event shape observables the 
QCD predictions were obtained by numerical integration of the QCD matrix
elements using the program EVENT2~\cite{event2}. 
A  $\chi^{2}$ value was minimized with respect to \asmz\ for each 
moment $\momn{y}{n}$ separately.  
The fit results are shown in figure~\ref{alphas}. 
The fit \momone{\mh} did not converge and therefore 
no result is shown.  We
observed values of \chisqd\ of $\mathcal{O}(1)$; the fitted QCD 
predictions including the running of \as\ were thus consistent with our
data.  However, we fond that for \momn{(1-T)}{n}, \momn{\cp}{n} and
\momn{\bt}{n} the fitted values of \asmz\ increased steeply with the
order $n$ of the moment used.  
This effect was in clear correlation with increasing ratio of the NLO
and LO coefficient with moment $n$ for \momn{(1-T)}{n},
\momn{\cp}{n} and \momn{\bt}{n}.
We considered only those results for which the NLO term was less than 
half the LO term, namely \momone{\thr}, \momone{\cp}, \momone{\bt}, \momn{\bw}{n} and
\momn{(\ytwothree)}{n}, $n=1,\ldots,5$ and \momn{\mh}{n},
$n=2,\ldots,5$; i.e.\ results from 17 observables in total.  The
purpose of this requirement was to select observables with an
apparently converging perturbative prediction. The statistical correlations between the
17 results were determined using Monte Carlo simulation at the
hadron level.
The result of the combination was
\begin{displaymath}
 \resultmom\;,
\end{displaymath}
above but still consistent with the world average value of
$\asmz=0.1182\pm0.0027$ ~\cite{bethke04}.  
Combining only the fit results from \momone{\thr}, \momone{\cp}, 
\momone{\bt}, \momone{\bw}, \momone{\ytwothree} and \momn{\mh}{2} yields a 
value of
\begin{displaymath}
\ensuremath{\asmz=0.1239\pm0.0001\stat\pm0.0008\expt\pm0.0009\had\pm0.0059\theo} \;.
\end{displaymath}
The slightly smaller error of \as\ reflects the fact that the lower
order moments were less sensitive to the multi-jet region of the 
event shape distributions. This leads to a smaller systematic uncertainty.
\subsection{Four-Jet Rate}
Besides event shape observables jet rates reflect the parton structure
of the event. The four-jet rate $R_{4}(\ycut)$ is the
fraction of four-jet events as a function of \ycut, the jet-resolution
parameter determined in the Durham scheme. 
In QCD the fraction of four-jet events $R_4$ is predicted in
NLO as a function of the strong coupling \as.
The prediction was calculated by the
program~\debr~\cite{nagy98b} combined with an all order resummation 
using the ``modified R-matching'' scheme. 
In the case of jet rates a single event usually contributed 
to several \ycut\ points in the
four-jet rate distribution and for this reason the data points were
correlated. The complete covariance matrix was determined and used 
in the $\chi^{2}$ fit for the extraction of \as. 
The $\chi^{2}$ value was minimized with respect to \as\ for each center-of-mass energy separately. The fit ranges covered the decreasing
parts of the distributions at large \ycut, where the perturbative QCD
predictions were able to adequately describe the data corrected for
hadronization.  
We found that the fit result from the 14~GeV data had large
hadronization and experimental uncertainties because the corresponding
corrections were large and not well known at this energy.  We therefore
chose not include this result in the combination.  The result of
the combination using all results with $\rs\ge 22$~GeV was
\begin{displaymath}
 \resultjr\;,
\end{displaymath}
consistent with the world average value of $\asmz=0.1182\pm0.0027$
~\cite{bethke04}.  
The results at each energy point are shown in figure~\ref{alphas}
and compared with the predicted running of \as\ based on the world
average value.  
\section{Summary}
In this note we present preliminary measurements of the 
four-jet rate and the first five moments of 
event shape observables at center-of-mass energies between 14 
and 44~GeV using data of the JADE experiment.  
The value of the strong coupling is determined to be 
\resultjrtot\ using the four-jet rate and 
\resultmomtot\ using moments of event
shape observables.

\end{document}